# Biological Computing Fundamentals and Futures

Balaji Akula & James Cusick
Wolters Kluwer Corporate Legal Services
New York, NY
*{balaji.akula; james.cusick}@wolterskluwer.com*

## Abstract

*The fields of computing and biology have begun to cross paths in new ways. In this paper a review of the current research in biological computing is presented. Fundamental concepts are introduced and these foundational elements are explored to discuss the possibilities of a new computing paradigm. We assume the reader to possess a basic knowledge of Biology and Computer Science.*

## Introduction

It is easy to miss nature's influence and subsequent impact on living forms. This applies to our day to day activities as well. Humans use a variety of gadgets and gizmos without realizing that the gadget could be working on a pattern already patented and perfected by Mother Nature. Computers and software are no exception. The last few decades have ushered in the age of computers. Electronics have invaded all walks of life and we depend on electronics to accomplish most of our day to day activities. As predicted by Dr. Gordon E. Moore, modern day electronics has progressed with miniaturization of electronic components. According to Dr. Moore, the miniaturization of integrated electronics will continue to be bettered once every 12 – 18 months with a reduction in cost (Moore, 1965). True to his prediction modern day chips have up to 1 million transistors per $mm^2$. However as with other things, miniaturization cannot continue forever, the laws of nature and in particular physics will soon catch up to impose a limit on the silicon chip. Such limitation will not prevent us from progression. The route is clear but the ways to reach it may be unusual. Imagine having billions of Deoxyribonucleic (DNA) acids instead of silicon chips powering the computer. The fact that silicon chips will even be replaced will be anathema to some but we are well on our way for some surprises. Hence it is imperative that software engineers have an understanding even if it just includes the basics of microorganisms and how they will impact computing.

Our fascination and its logical conclusion, which is reflected in this paper, is due to the behavioral similarity between microorganisms (DNA) and computers. As soon as you understand what microorganisms can do, then relating that to a computer or program that runs on a computer becomes easy. Much like microorganisms, computers have evolved over a period of time. However time will tell if DNA will indeed play a prominent role in their march to future glory. It is our endeavor to shed light on biological computing thru a lay person's eyes.

## Concept

This paper talks about how two diverse systems, biology and computers are brought together to take mankind into the future. A basic understanding of the lowest unit (Deoxyribonucleic acid - DNA) of life will help. People should not imagine that DNA will replace the CPU in biological computing. In our opinion such a scenario is at least two decades or more away from reality. As like other inventions one can safely anticipate or expect baby steps in this direction before conceiving bigger pictures. Although not exceeding a few microns in size, the DNA molecule has a number of tricks that will be useful for biological computing. One of them is the ability to generate proteins. Once programmed, by altering the cell by chemical or changing the environment the reprogrammed cell does its job to near perfection as per the changed environment Another trick that may be useful is the ability of DNA to make exact copies of itself. Imagine the advantage of having such molecules programmed for different purposes and its impact on applied sciences like medicine, agriculture, and various industries, in fact such molecules act like micro computers. There is no clear road map for this programmable feature to be taken advantage of to eventually replace the CPU. In essence, Biological computing is about harnessing the enormous potential of the DNA to the benefit of mankind by manipulating the DNA.

Having laid down the concept and to provide clarity to better understand and appreciate biological computing we are providing a brief introduction to DNA. We will also provide the similarities between DNA and the computer; briefly provide information on current research and finally touch upon trends, impact and future prospects.

## *Deoxyribonucleic Acid (DNA)*

This section provides a summary of DNA. This detailed information can be found in any biology book but is condensed here to set up this discussion of bio-computing. The essence of life is enclosed in a 20 micron long substance called DNA. The structure of DNA (Figure 1) was first identified by Watson and Crick (1953). The earliest discovery of DNA was by Swiss physician Fritz Miescher in cell nuclei as early as 1868. According to the Watson–Crick model, the DNA molecule consists of two polymer chains. Each chain comprises four types of residues (bases) – namely A (adenyl), G (guanyl), T (thymidyl), and C (cytidyl). The sequence of bases in one chain may be entirely arbitrary, but the sequences in both chains are strongly interconnected because of the complementary principle so that:

> A is always opposite T;
> T is always opposite A;
> G is always opposite C;
> C is always opposite G.

DNA was recognized as the most important molecule of living nature. In living organisms, DNA does not usually exist as a single molecule, but instead as a tightly-associated pair of molecules. These two long strands entwine like vines, in the shape of a double helix. The nucleotide repeats (structural units of DNA, Figure 1) contain both the segment of the backbone of the molecule, which holds the chain together, and a base, which interacts with the other DNA strand in the helix. In general, a base linked to a sugar is called a nucleoside and a base linked to a sugar and one or more phosphate groups is called a nucleotide. If multiple nucleotides are linked together, as in DNA, this polymer is called a polynucleotide (Frank-Kamenetskii, 1997). At the time of discovery of the structure by Watson-Crick, it was a great step for mankind in the field of biology but very difficult to have dreamt that half a century later it will also help mankind in another field – computing.

## *Biological Computing Simplified*

The above brief definition about DNA may be Latin and Greek to pure computer engineers. We hope to change that via Figure 2 given below. In the figure, under Laboratory conditions it is possible to take part of a DNA molecule and engineer it to reproduce a particular protein (the end product of a successful DNA transcription is a protein).

Computers use registers to flip the binary between 1 and 0. In microorganisms the same "Registers" and flip-flop occurs but at the DNA level. In the above example you can imagine Adenine, Cytosine, Guanine, and Thymine are the registers that are involved in protein synthesis. Any change to this structure or inhibition of the normal protein synthesis by changes in environment results in a completely new product; worse in some cases if no product is created. This is the whole idea of using DNA (refer to Concept section) in biological computing. As you can see the comparison between DNA and the computer is as close as one can imagine.

DNA is ubiquitous in life forms and is self contained. Its intelligence and ability to adapt to changing conditions far surpasses anything and everything one could imagine. A double stranded DNA within a single cell is fully self contained. It works with clockwork precision, has the ability to repair itself; provide backup; create new patterns; select the best for its survival. Most complex computers exhibit the above in one way or the other. As a DNA has to survive in nature, only the fittest survive and hence the ability to adapt to changing environment. However the same environment (extreme heat, chemicals, Ultra Violet rays, etc.) can sometimes causes changes to DNA that may make them loose some of their magic and in some cases can be catastrophic. In real life, the DNA is intelligent enough to recover from catastrophic failures. There are many tools that it carries to successfully replicate thereby passing on the important traits to its progeny. A few of them include redundancy, self recovery by protein synthesis/translation, and ability to shut down malfunctioning parts of DNA.

Compare this with a computer and the software that runs the computer. Even a pure software engineer will now be able to link the computer to the DNA. In fact I would go as far to say that what we know in computer jargon as "Primer", "Reusability", etc., has been in existence since time immemorial in the DNA molecule. Microsoft had in fact coined the terminology "DNA" in the late 1990's to market their Distributed networking solutions (since then Microsoft has dropped it for whatever reasons) and one can safely assume that they had borrowed it from biology. Table 1 below compares a DNA with a modern day computer.

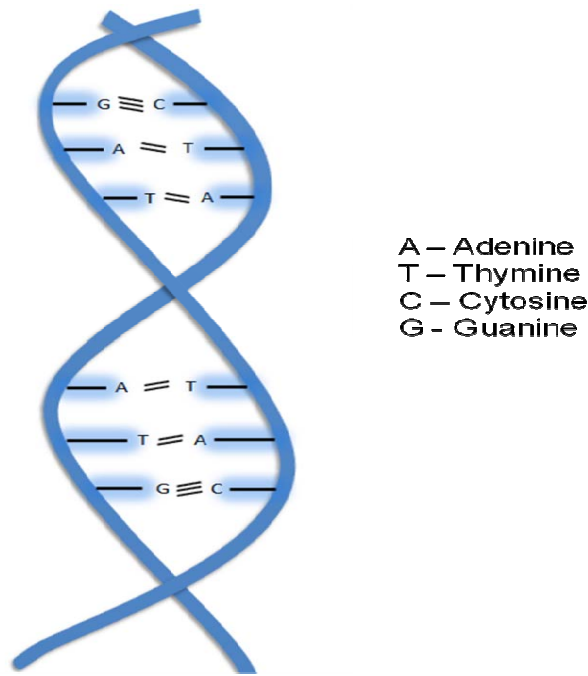

*Figure 1:* *Structure of Single strand of DNA (only a portion shown)*

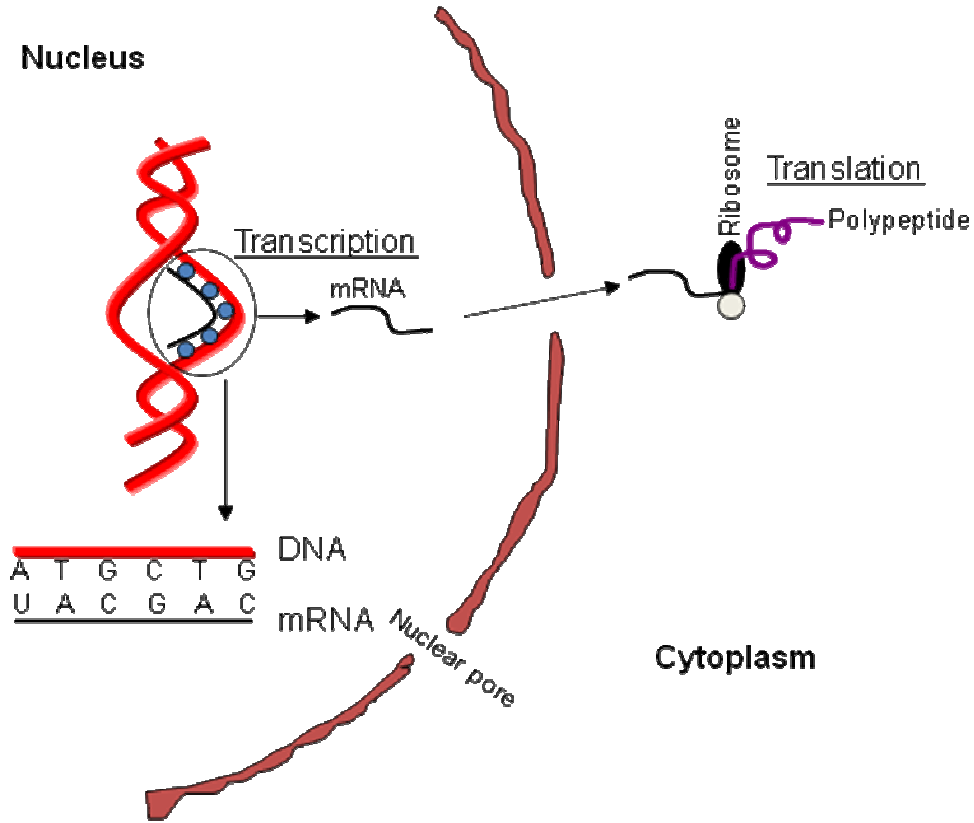

*Figure 2:* *Simplified diagram of Protein synthesis*

We would like to touch upon a few of the points mentioned in the above comparison table to highlight the benefits of taking biological computing to its next step, which is to make it a reality. The ability to store billions of data is an important feature of the DNA and hence to biological computing. While DNA can be measured in nano grams, the silicon chip is far behind when it comes to storage capacity. A single gram of DNA can store as much information as 1 trillion audio CDs (Fulk, 2003). This offers storage possibilities previously unheard of and at the same time businesses can reduce the cost of storage and plough investments into other areas. While we are all familiar with Von Neumann's sequential architecture which has stood the test of time, the fact that we could have millions of DNA molecules in a small vial allows us to think of massive parallel computing when using microorganisms. Parallel processing using DNA can achieve speeds that man could not have imagined. For comparison, the fastest supercomputers can perform around $10^{12}$ operations per second, but even current results with DNA computing has produced levels of $10^{14}$ operations per second or one hundred times faster. Experts believe that it should be possible to produce massively parallel processing in biological computers at a level of $10^{17}$ operations per second or more, or a level that silicon-based computers will never be able to match (Fulk, 2003).

| DNA | Computer / Programs |
|---|---|
| Fully Autonomous | Self contained to a great degree – dependent on peripherals, power etc. |
| Has inbuilt redundancy | Depending on need has redundancy built inside |
| Ability to recover from failure is remarkable – redundancy, shut down etc., | Depending on need. Redundancy, backups, disks, additional power sources etc help systems to recover. |
| Can adapt to environment | Not available |
| Store billions of pieces of information due to their size | Limited by technology |
| Can reproduce information with precision | Can reproduce information with precision – Garbage in and Garbage out |
| Can be manipulated by external stimulus – chemicals, heat, etc., | Can be manipulated by external stimulus – mouse, external commands etc |
| Impacted by environment changes | Less Impacted by environment changes |
| No toxic byproduct is generated | Composed of Toxic products and generates lot of heat |
| Energy Efficient | Less Energy Efficient (generates heat) |

*Table 1:* *Comparison between DNA and current computers*

Second, in the case of DNA computing, the biological reactions involved produce very little heat, wasting far less energy in the process. This allows for these computing processes to be up to one billion times as energy efficient as their electronic counterparts.

Third, the components of a computer composed of DNA as the primary unit is non toxic when compared to the current systems which is highly toxic due to use of chemicals and other materials that are not easily degradable. Not only is the material toxic but in some cases production of such materials also results in toxic byproducts. The damage of such toxic materials to the environment is unimaginable and the cost to clean up is also high.

Lastly, DNA has the inbuilt ability to repair itself in case of any impact to its functioning. This type of self-healing is not possible in a hardware based computer. It may sound a bit like an H.G Wells story, but imagine a computer that does not break down after a few years in operation and one that does not require a hardware upgrade? The benefits of moving towards biological computing appears immense.

## *Current Research*

Before embarking on this paper we did some research to find out where the world is in terms of biological computing. As one would expect we see a lot of baby steps being taken in this field. Part of the reason is because software engineers need to first understand Biological sciences. It is a radically different field where there is no easy way to debug; to fix and run a program. Take for example the "Genetic Circuit", worked on by Michael Elowitz and his team (Garfinkel, 2000). The circuit consists of four genes engineered into a bacterium. Three of them work together to turn the fourth, which encodes for a fluorescent protein, on and off. Although this circuit is a remarkable achievement, it doesn't keep great time—the span between tick and tock ranges anywhere from 120 minutes to 200 minutes. And with each clock running separately in each of many bacteria, coordination is a problem: watch one bacterium under a microscope and you'll see regular intervals of glowing and dimness as the gene for the fluorescent protein is turned on and off, but put a mass of the bacteria together and they will all be out of sync. This is a big first attempt and we have many miles to go (Garfinkel, 2000).

Another interesting work with a name that almost rhymes like a software object is being carried out by James J. Collins at Boston University. The main focus is on "Genetic Applets". Similar to what a Java applet is and does the genetic applet is modeled on the same lines, i.e. programmatically altered to perform one or more functions repeatedly with perfection (Garfinkel, 2000).

One might wonder how such DNA molecules that are programmed for one or a few functions can one day replace the CPU. To answer this one must look into the work that is carried out by Dr. Thomas F. Knight. His team has forayed into what is known as amorphous computing. Knight's lab is working on techniques to exchange data between cells and between cells and large scale computers as communication between components is a fundamental functionality of computers. The concept of bioluminescence is used for this purpose. Needless to say all of the techniques involve splicing and dicing of genetic materials which is nothing but the DNA.

## *Trends, Challenges & Future Prospects*

It is clear that scientist and various teams have been working to realize the huge potential of the DNA molecule. James J. Collins and his team have gone to the extent of enabling communication between the molecules and a computer. Knowingly or unknowingly biology has been the inspiration for computers to a great extent. The similarities are too many to think otherwise. So it is time for harnessing the power of DNA using computers as the inspiration.

While we live in the age of computers, biological computing is slowly gaining prominence but without much fanfare. True, biological computing has played a big role in modern medicine and will continue to do so, but to see a computer being solely powered by microorganisms/DNA is far away. We feel that we are not even close enough to say that the next years will see the dawn of biological computing where CPU is replaced by DNA. Some of the challenges that stare us in our face to eventually replace silicon chips with DNA include:

a) Ability to control the DNA.
b) How to make the various altered DNA's to communicate with each other.
c) Can the programmed DNA or microorganism go wrong?
d) Can it impact health?

Maybe the above may not be an issue at all but still they need to be answered. For all those hard core computer professionals who are wedded to silicon chips it is time to look at the future and prepare for the next big thing in computers.

The future for biological computing is bright. Already some of the medical/industrial products like Vaccines, Insulin (for diabetes treatment) are benefiting from this research. Most of the design/patterns coming out of various software companies have already been in existence in nature (DNA) and all we need to do to effectively use the DNA is to reverse engineer, understand the inner workings and make it fit to work to our requirements. The advent and gaining popularity of Nano technology offer more avenues to use DNA. Under laboratory conditions, DNA self-assembly has been demonstrated successfully, simple patterns (e.g., alternating bands, or the encoding of a binary string) that are visible through microscopy has been used successfully for simple computations such as counting, XOR, and addition (Wooley and Lin, 2005).

## Conclusion

Biological computing is a young field which attempts to extract computing power from the collective action of large numbers of biological molecules. In our opinion the CPU being replaced by biological molecules remains in the far future. However if one can imagine such a scenario then it is safe to imagine or think of a biological computer as a massively parallel machine where each processor consists of a single biological macromolecule. By employing extremely large numbers of such macromolecules in parallel, one can hope to solve computational problems more quickly than the fastest conventional supercomputers. To many pure software professionals this may be far-fetched. A good compromise could be a hybrid system. A part of the system can be made of biological and the other using current or new hardware that may become available. This would give us the combined benefit of both systems. Companies and scientist that are involved in the biological computing work need to take care of legal, moral regulations. Maybe it is time to overcome Moore's law as the rate of doubling has slowed down (Mathews, 2006). We need to think of computing in a radically different way and who knows if in the near future we will be tackling real viruses instead of the electronic virus.